# Implementation of A Nanosecond Time-resolved APD Detector System for NRS Experiment in HEPS-TF

LI Zhen-jie[a]; MA Yi-chao[c]; LI Qiu-ju[a]; LIU Peng[a]; CHANG Jin-fan[b]; ZHOU Yang-fan[a*]

[a] Beijing Synchrotron Radiation Facility, Institute of High Energy Physics, Chinese Academy of Sciences

[b] State Key Laboratory of Particle Detection and Electronics, Institute of High Energy Physics, Chinese Academy of Sciences

[c] Shaanxi University of Science & Technology

*Correspondence e-mail: zhouyf@ihep.ac.cn

**Abstract**：A nanosecond time-resolved APD detector system is implemented for Nuclear Resonance Scattering (NRS) experiments in High Energy Photon Source-Test Facility (HEPS-TF) project of China. The detector system consists of three parts: the APD sensors, the fast preamplifiers and the TDC readout electronics. To improve the reception solid angle and the quantum efficiency, the C30703FH APDs (fabricated by Excelitas) are used as the sensors of the detectors. The C30703FH has an effective light-sensitive area of $10 \times 10$ mm$^2$ and an absorption layer thickness of 110 μm. The fast preamplifier with gain of 59 dB and bandwidth of 2 GHz is designed to readout the weak signal outputted by the C30703FH APD. The detector system can work in single photon measurement mode because the preamplifier increases the signal-to-noise ratio. Moreover, the TDC is realized by FPGA multiphase method with a resolution bin of 1ns. The arrival time of all scattering events between two start triggers can be recorded by the FPGA TDC. In the X-ray energy of 14.4 keV, the time resolution (FWHM) of the developed detector (APD sensor + fast amplifier) is 0.86 ns, and the whole detector system (APD sensors + fast amplifiers + TDC readout electronics) achieves a time resolution of 1.4 ns.

**Key words**: HEPS-TF, Time-resolved, APD, fast amplifier, TDC

**PACS:** 07.85.Qe, 29.20.dk, 29.40.Wk

## 1. Introduction

The silicon avalanche-photodiodes (APD) detector system has a number of advantages such as high counting rate, large dynamic range, and nanosecond or faster time resolution. It is widely used in Nuclear Resonance Scattering (NRS) experiment with synchrotron radiation [1]. In the NRS experiment [2, 3], a strong scattering due to electrons occurs promptly (on a ps time scale) after the incident pulse, but the scattering due to the resonant nuclear excitation is delayed (on a 10's of ns time scale) by the finite lifetime of the excited state. The detailed time structure of the electronic scattering and the nuclear scattering can be obtained by nanosecond time-resolved APD detector system. Therefore, we can separate the nuclear scattering events from the electronic scattering events.

The High Energy Photon Source-Test Facility (HEPS-TF) is a pre-research project for the future High Energy Photon Source in China. One of its subprojects is the High Resolution Monochromator, which is mainly designed for the NRS experiment. The APD detector system is an important part of the High Resolution Monochromator subproject. It will be an essential equipment for NRS experiment in future High Energy Photon Source of China. Furthermore, the detector system is also can be used to carry out the laser pump / X-ray probe experiment, test the bunch structure, and measure bunch-purity at Beijing Synchrotron Radiation Facility (BSRF).

Some time-resolved APD detector systems have been reported by several synchrotron sources (ESRF, Spring-8, KEK, APS, and so on) [4-8]. The APD detector systems reported by Spring-8 or KEK are based primarily on APD sensors from Hamamatsu. And the APD sensors manufactured by EG&G are usually adopted by ESRF or APS to design the APD detector systems. Those detector systems have a time resolution of 0.1 ns ~ 1.7 ns. However, the time measurement electronics of those traditional detector systems is based on the NIM-modules, which can just record the arrival time of the first delayed nuclear scattering event between two start triggers. Moreover, the NIM-module readout electronics has a low integration level and is significantly unpractical for multiple APD detector system or APD array detector system in the future.

This paper implements a highly integrated APD detector system for NRS experiment. The implemented detector system has four APD sensors, four fast preamplifiers, and a time-digital-convert (TDC) readout electronics. The reception solid angle and the quantum efficiency is increased by using the large effective light-sensitive area APD sensor with thick absorption layer. In order to obtain a sufficient signal-to-noise ratio to realize single photon measurement mode, the fast preamplifier with three cascade amplification stages is designed for readout the weak signal from APD sensor. The TDC readout electronics integrates the NIM functions into one board so that the integration of the entire detector system is greatly improved. The time measurement is based on field programmable gate array (FPGA) TDC, which can record the arrival time of all scattering events between two start triggers. The detail design of the developed detector system will be discussed in the following sections. The measurement results is also presented to demonstrate the capability of the detector system.

**2. Nanosecond time-resolved APD detector system design**

The specifications of the designed APD detector system are defined to accomplish the $^{57}$Fe NRS experiment. The excited state energy of $^{57}$Fe is 14.4 keV, while the excited state lifetime is 141 ns. This implies a quantum efficiency of the APD detector system must be larger than 10 % at 14.4 keV, and the count rate should be larger than $10^6$ /s. Considering the detector must recover from an electronic scattering event quickly to see a nuclear scattering event a few 10's ns later, the output pulse width of the preamplifier < 30 ns is reasonable. In most experiments the nuclear scattering events rate may as low as 0.1 Hz, so the count noise rate of the detector system should be lower than 0.01 Hz. Besides, the time resolution < 2 ns is also needed to capture the arrival time of scattering event accurately. The design specifications are summarized in Table 1. The simultaneous requirement of high quantum efficiency, high count rate, fast recovery capability, nanosecond time resolution and very low noise background rate set a new challenge in the detector system design.

**Table1** Design specifications of the implemented APD detector system

| Parameter | Design specifications |
| --- | --- |
| Quantum efficiency | > 10 % |
| Count rate | > 1 MHz |
| Output pulse width of the pre-amplifier | < 30 ns |
| Count noise rate | < 0.01 Hz |
| Time resolution | < 2 ns |

Fig.1 shows the implemented APD detector system structure. The detector system is composed by four detectors and a TDC readout electronics. Each detector consists of an APD sensor and a fast preamplifier. The TDC readout electronics include four readout channels. Each readout channel can finish the functions of signal inversion, discrimination and TDC. The four detectors connect the TDC readout electronics with high bandwidth cables. The time data, obtained by the TDC readout electronics, is transmitted to DAQ via a single optical fiber.

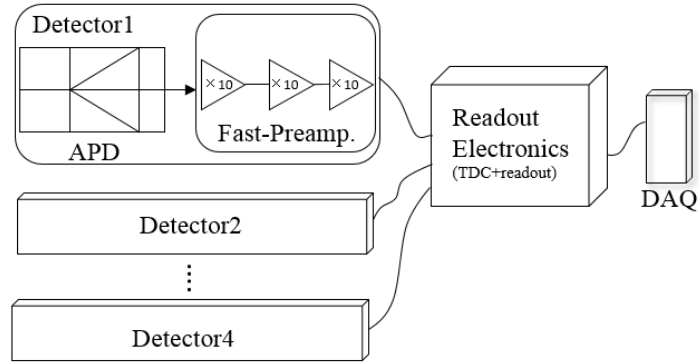

Fig. 1. Structure of the detector system

**2.1 Si-APD sensor**

The APD sensor used in this detector system is C30703FH which fabricated by Excelitas. The sensor is designed on a 120um thick silicon wafer with reach-through structure [9]. The avalanche amplification region located in the back of the device is very narrow, so the device presents, at normal incidence, an absorption layer thickness about 110 μm. This absorption thickness makes the sensor quantum efficiency can reach 25% at the energy of 14.4keV. The sensitive area of this sensor is available in $10 \times 10$ mm$^2$, which provides a sufficient reception solid angle for X-ray detection.

**2.2 Fast preamplifier**

The current signals from the APD are processed with a high bandwidth (2 GHz), high gain (59 dB) preamplifier. The schematic diagram of the fast preamplifier circuit is shown as fig. 2. The preamplifier consists of three commercial amplifier chips and two π type resistance circuits. These amplifier chips are cascaded in three stages to get a gain of about 60dB. To remit the oscillation, which often appears in the high bandwidth and high gain circuit, the π type resistance circuits are added between two amplifier chips.

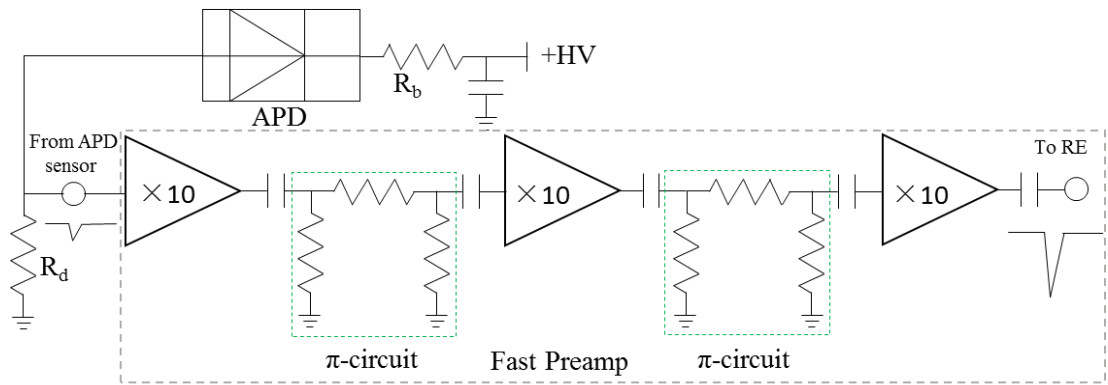

Fig. 2. Schematic of the fast preamplifier. $R_b$ (470kOhm) limits the bias current and $R_d$ (27kOhm) is closed to the bias circuit. The pulse of the fast signal from APD goes through capacitance to the first stage amplifier and to the ground. This is the low impedance (50Ohm) path way by the signal. The circuit is mounted on the circuit board with a copper ground plane.

The input impedance of the π type resistance circuit should be matched with the output impedance of the signal, and the output impedance of the π type resistance circuit should be matched with the load. In this design, the impendence of the circuit is 50 Ω, then the input impedance of the π type resistance circuit $Z_{in}$ should be

$$Z_{in} = R_2 \| (R_1 + R_3 \| 50\Omega) = 50\Omega$$

Suppose the gain attenuation for the π type resistance circuit is N dB, then

$$\frac{R_3}{R_1 + R_3} = 10^{-\frac{N}{20}}$$

In this paper, the values of the resistances are $R_1 = 10\ \Omega$, $R_2 = R_3 = 300\ \Omega$. The attenuation for one π type resistance circuit is 0.28 dB. Since the gain of the commercial amplifier chip is 20 dB, the preamplifier has a gain of 59 dB. The main function of the π type resistance circuits is to separate the commercial amplifier chips and make sure the preamplifier operate without oscillation.

The fast preamplifier is first designed and fabricated on a printed circuit board. The preamplifier board is covered by an aluminium box to shield the environment electromagnetic radiation. Furthermore, the high voltage circuit of the APD sensor is isolated from the fast preamplifier by an aluminium wall in the box. Therefore, the high voltage circuit will not induce an oscillation to the fast preamplifier. Through the above techniques a stable preamplifier with high bandwidth and high gain is implemented.

**2.3 TDC readout electronics**

Fig. 3 shows the logic diagram of the time measurement for NRS experiment. The timing signal is the RF signal from the storage ring and it acts as the start signal of the TDC. Each timing signal correspond with each X-ray bunch. The arrival time, defined as the difference time between the timing signal and the prompt signal or the delay signal, is measured by TDC. The prompt and delay signals act as the stop flags of the TDC. In this paper, the TDC readout electronics can measure one or more scattering events after the timing signal.

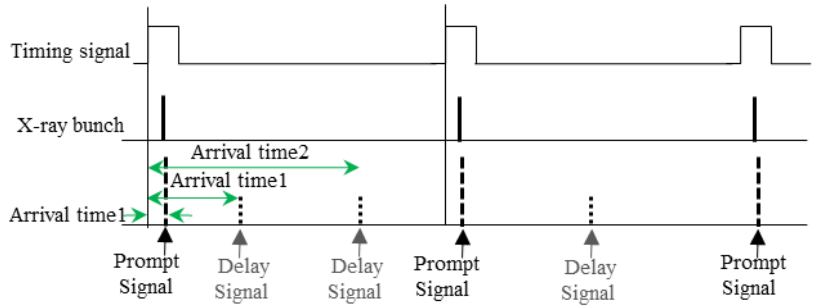

Fig. 3. Logic diagram of the time measurement for NRS experiment

The TDC readout electronics includes four readout channels. Fig. 4 shows the structure of one channel. Signals from the fast preamplifier are modulated by the next stage (-A1), which can magnify the signals several times and change the polarity of the signals from negative to positive, so that the signals could be discriminated by the discriminator (TH). Moreover the digital to analog converter (DAC) is added to adjust the discrimination threshold. Then the interval time between the start signal (timing signal from the storage ring) and the stop signal (output signal of the discriminator which represents the scattering signal) is measured by the FPGA TDC.

This TDC uses a multi-phase technology and realized in an FPGA chip [10]. It formed by the stop-edge detector, the start-edge detector (shared by four channels), the coarse-time counter, the time-calculate unite, and the phase lock loop (PLL) (also shared by four channels). The PLL creates a 250 MHz quadrature clock and a 125 MHz working clock. The quadrature clock has four phases with 0-, 90-, 180-, and 270-degrees. Each phase of the quadrature clock operates at 250 MHz, essentially creating a clock that operates at 1 GHz. The stop-edge detector, clocked by the 0-, 90-, 180-, and 270- phases of the quadrature clock, detects the rising edge of the stop signal and records the stop hit-time in every period of 8 ns (two quadrature clock cycles). Thanks to the quadrature clock can equivalent to a 1 GHz clock, the resolution bin size of the stop-edge detector is 1ns. The start-edge detector records the start hit-time using the same working principle as the stop-edge detector. The coarse-time counter counts at a count clock of 125 MHz, with a period of 8 ns. This counter is reset to zero by every rising edge of the start signal and resumes counting at the first rising edge of the count clock after the start signal. When the rising edge of the stop signal arrives, the coarse-time data recorded by the coarse-time counter is sent to the time-calculate unite. At the same time, the time calculation unit calculates the arrival time (the interval time between start signal and stop signal) by combining the start hit-time, stop hit-time and the coarse-time. The timing diagram of the TDC is simply illustrated in Fig.5, and the arrival time = start hit-time + coarse-time + stop hit-time, with a fixed time-stamp resolution of 1 ns per bin. Then, the data of the arrival time are output through the first-in first-out (FIFO) buffer and the user datagram protocol (UDP) to the data acquisition (DAQ) system.

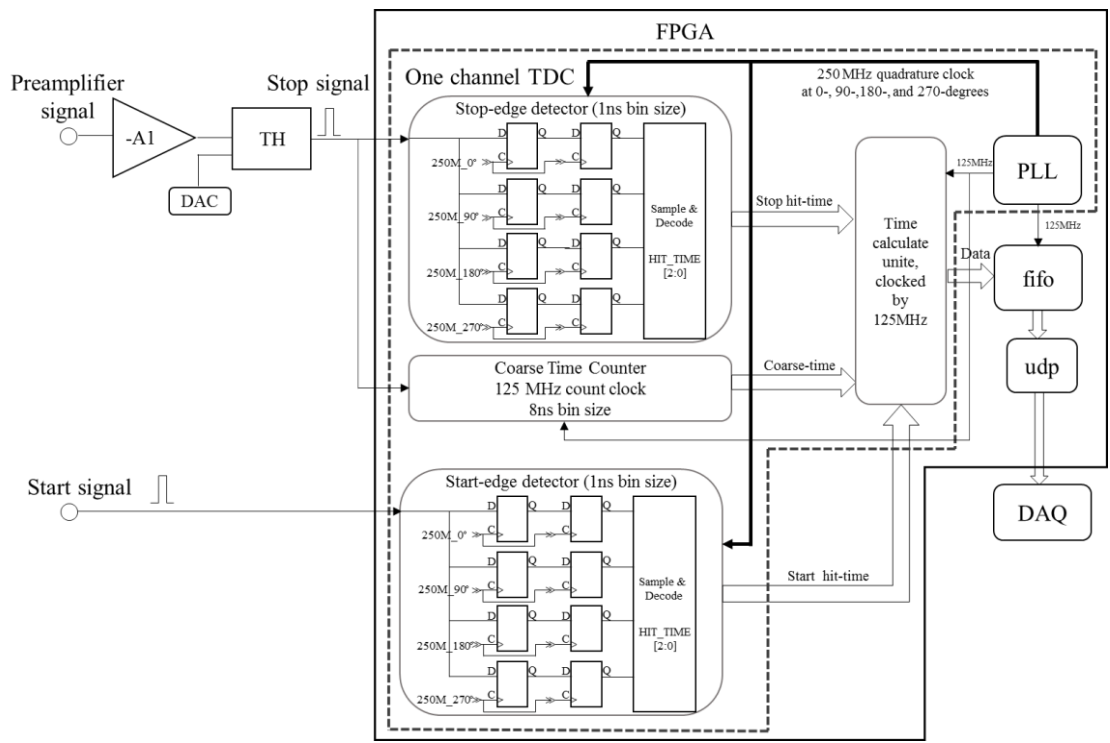

Fig.4 Structure for one channel of the TDC readout electronics

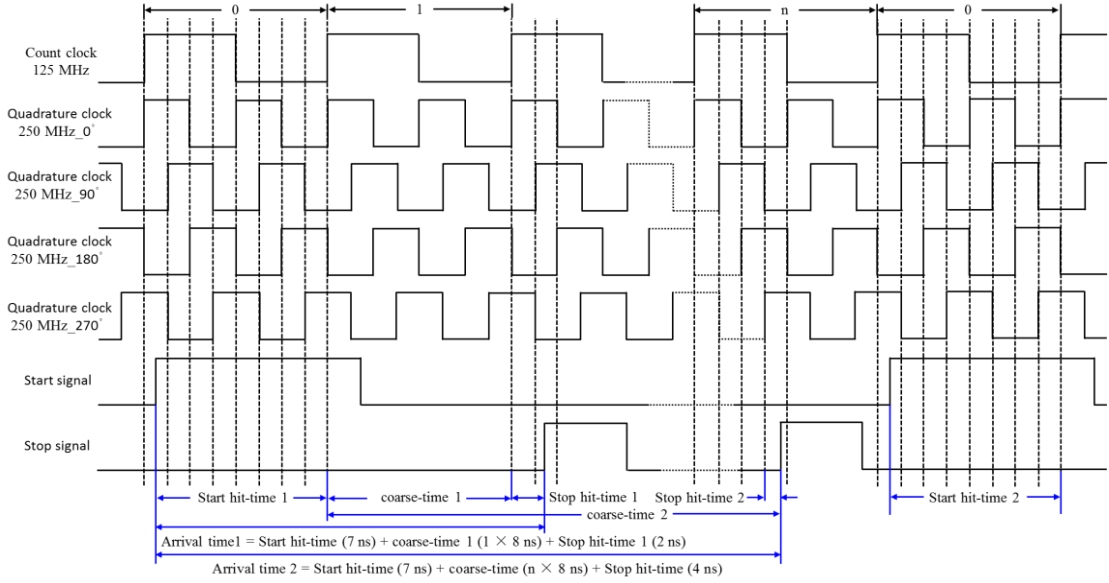

Fig. 5 Timing diagram of the TDC

## 3. Performance

### 3.1 APD detector performance

The APD sensor and the fast preamplifier are installed in an aluminium case and tested as an X-ray detector at 1W2B experiment station of BSRF. The output pulse from the detector is shown in Fig. 6 (a), and the pulse width is about 25 ns. In order to estimate the count rate, the APD detector linearity at high count rate was investigated. The measured counting rate versus the input X-ray intensity (monitored by the ion chamber) is shown in Fig. 6 (b). From the result, one can see the detector has a good linearity below the count rate of 8 MHz. Considering the noise rate is as low as 0.01 Hz, the dynamic range of the detector is larger than $8 \times 10^8$.

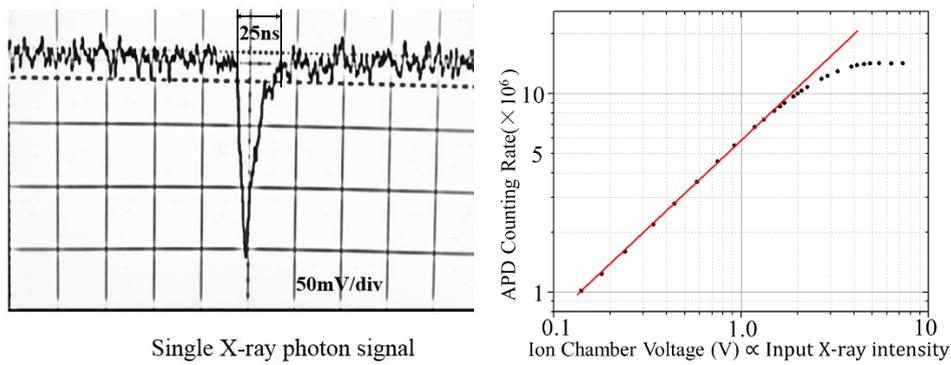

Fig. 6. (a) Single X-ray photon signal pulse, (b) Counting rate vs input X-ray intensity

The time resolution of the designed APD detector is also investigated by measuring the arrival time of the X-ray photon after the locked RF signal from the accelerator storage ring. The FWHM time resolution of the detector is 860ps (@14.4 keV), as shown in Fig. 7. This result indicates that the time resolution of the subsequent readout electronics including the TDCs wouldn't need to design less than 860 ps. Moreover, almost all NRS experiments require a time resolution at a few ns level. Therefore, the TDC readout electronics in our project is designed to a resolution bin of 1ns.

The designed detector also has been used in APS 3-ID beamline for $^{57}$Fe NRS experiment, and the obtained experiment results are shown in Fig. 8. The time spectrum of the two samples is measured in Fig. 8 (a), and the corresponding mössbauer spectrum is recorded in Fig. 8 (b). These results demonstrate that the designed detector can meet the requirements in NRS experiment perfectly. Two of the designed APD detectors are still using in the APS 3-ID beamline, and have been highly appreciated by experts of APS.

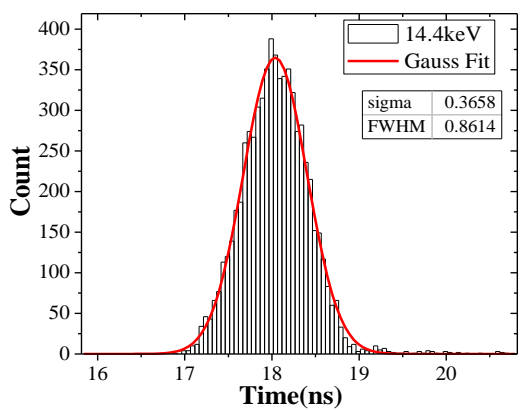

Fig. 7. Time resolution of the APD detector

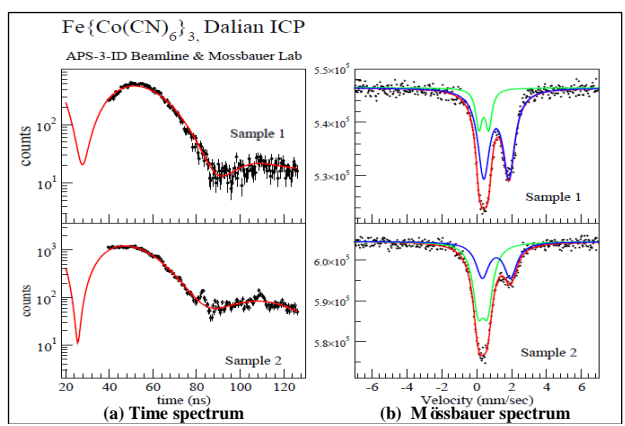

Fig. 8. $^{57}$Fe NRS experiment results with the designed detector

**3.2 Time resolution of the detector system**

The designed APD detector system is also tested at 1W2B experiment station of BSRF. In the test experiment, the RF signal from the accelerator storage ring serves as a start signal of the TDC and the X-ray photon signals are the stop triggers for the TDC. Fig. 9 shows the measured time structure of the beam bunch of BSRF by using the developed APD detector system. The beam bunch structure has a hybrid fill pattern in which a specific single bunch is filled with sufficient interval from other bunches. This measured bunch structure is consistent with the bunch current structure from the bunch current monitor, and proves the detector system can work in a healthy state. Fig. 10 shows the measured time resolution of the detector system at 14.4 keV. The time resolution (FWHM) is 1.4 ns which satisfies the design goal of < 2 ns. The measure results demonstrate the developed APD detector system has reached the design specifications for the NRS experiments in HEPS-TF project.

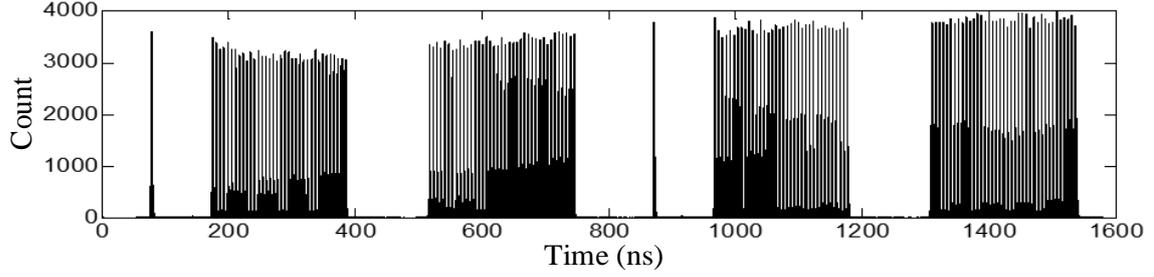

Fig. 9. Time structure of the beam bunch at BSRF

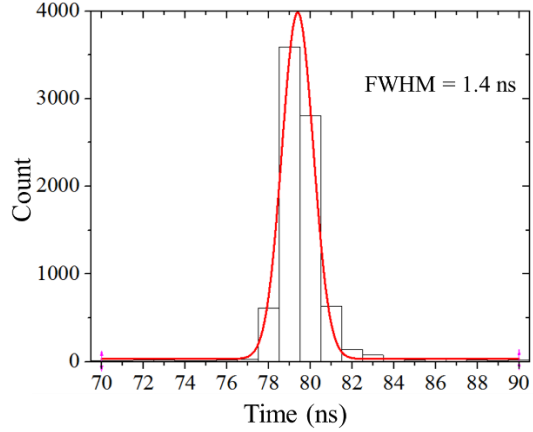

Fig.10 Time resolution of the APD detector system

## 4. Conclusions and outlook

In this paper, we have implemented a nanosecond time-resolved APD detector system for NRS experiments in HEPS-TF project. The detector system is formed by four APD sensors, four fast preamplifiers, and a TDC readout electronics. The effective light-sensitive area of $10 \times 10$ mm$^2$ and the quantum efficiency of 25% are achieved by utilizing the C30703FH APD as the detector sensor. The fast preamplifier is designed on the discrete components with gain of 59 dB and bandwidth of 2 GHz. While the TDC, with a resolution bin of 1ns, is realized by FPGA multiphase method, which can record the arrival time of all scattering events between two start triggers. The measured results indicate that the developed APD detector time resolution is 0.86 ns and the APD detector system time resolution is 1.4 ns. This novel APD detector system will greatly promote the development of the NRS experiment in future High Energy Photon Source of China.

**Acknowledgements**

This work is supported by the National Natural Science Foundation of China (Grant No. 11605227), the Research and Development Project for Scientific Research Conditions and Resources of Hubei Province of China (No. 2015BCE076), High Energy Photon Source-Test Facility Project, and the State Key Laboratory of Particle Detection and Electronics.